\documentclass[preprint,showpacs,showkeys,preprintnumbers,amsmath,amssymb]{revtex4}

\usepackage{graphicx}

\usepackage{dcolumn}
\usepackage{bm}
\usepackage{url}

\newcommand{\ud}{\mathrm{d}}
\newcommand{\ui}{\mathrm{i}}

\begin{document}

\preprint{RSI/Kumar-Bellan1}


\title{Heterodyne interferometer with unequal path lengths}

\author{Deepak Kumar}
\email[]{deeku@caltech.edu}
\homepage[]{www.its.caltech.edu/~deeku/}

\author{Paul M. Bellan}
\email[]{pbellan@caltech.edu}
\homepage[]{http://www.aph.caltech.edu/people/bellan_p.html}
\affiliation{Applied Physics, California Institute of Technology, Pasadena, CA 91125.}

\date{\today}

\begin{abstract}
Laser interferometry is an extensively used diagnostic for plasma experiments. Existing plasma interferometers are designed on the presumption that the scene and reference beam path lengths have to be equal, a requirement that is costly in both the number of optical components and the alignment complexity. It is shown here that having equal path lengths is not necessary - instead what is required is that the path length difference be an even multiple of the laser cavity length. This assertion has been verified in a heterodyne laser interferometer that measures typical line-average densities of $\sim 10^{21}/\textrm{m}^2$ with an error of $\sim10^{19}/\textrm{m}^2$.
\end{abstract}

\maketitle

\section{\label{sec:introduction}Introduction}
Laser interferometry is an unambiguous diagnostic tool for measuring line-integrated plasma density and so is extensively used. It is generally believed that the path lengths of the scene and reference beams of the interferometer must be equal and this requirement has been incorporated into existing interferometers\cite{Acedo:TJ2:2000,Baker:D3D:1978,Carlstrom:D3D:1988,Irby::1999,Yasunori:JT60U:1997,Irby:TJ2:2001}. This leads to complicated optics alignment procedures over long distances. The assumed reason for maintaining equal path lengths of the scene and reference beams is to ensure coherent phases of the two beams. The extent to which the phases of the beams are coherent is determined by the phase auto-correlation function of the laser.

The purpose of this paper is to show that having equal path lengths is unnecessary. The paper is organized as follows. Section \ref{sec:laser-phase-auto-correlation} shows that the phase auto-correlation function, a measure of the coherence, is a quasi-periodic function of the path length difference between the two beams. The phase auto-correlation function peaks at path length difference $\delta l = 0,\pm 2d,\pm 4d,\pm 6d,\ldots$, where $d$ is the cavity length of the laser. Interferometer design can be simplified by operating at a path length difference corresponding to one of these peaks. Section \ref{sec:interferometer-setup} describes the setup of an interferometer being used on a spheromak formation experiment at Caltech\cite{You:caltech:2005}. The interferometer uses a laser with a $25$ cm cavity length and operates at a path length difference of $\sim 8$ m. Typical results from this interferometer are presented in Section \ref{sec:results}. Section \ref{sec:conclusion} concludes the paper with a summary.
\section{\label{sec:laser-phase-auto-correlation}Laser Phase Auto-Correlation}
\subsection{\label{subsec:laser-frequency-spectrum}Frequency spectrum of the laser}
A gas laser contains an active medium within a resonating optical cavity bounded by mirrors on either end. The mirrors allow only those optical modes which traverse an integer number of half-wavelengths within the cavity. The frequencies of these optical modes are
\begin{equation}
\nu_q=q\frac{c}{2d} \qquad q=0,1,2\ldots,
\label{eq:resonator-discrete-frequencies}
\end{equation}
 where $c$ is the speed of light and $d$ is the distance between the cavity mirrors. These discrete frequencies are separated by $\nu_M=c/2d$. For a typical He-Ne gas laser with a cavity length of $d\sim 25$ cm, the modes are separated by $\nu_M \sim 600$ MHz.
\begin{figure}[p]
\includegraphics[width=0.45\textwidth]{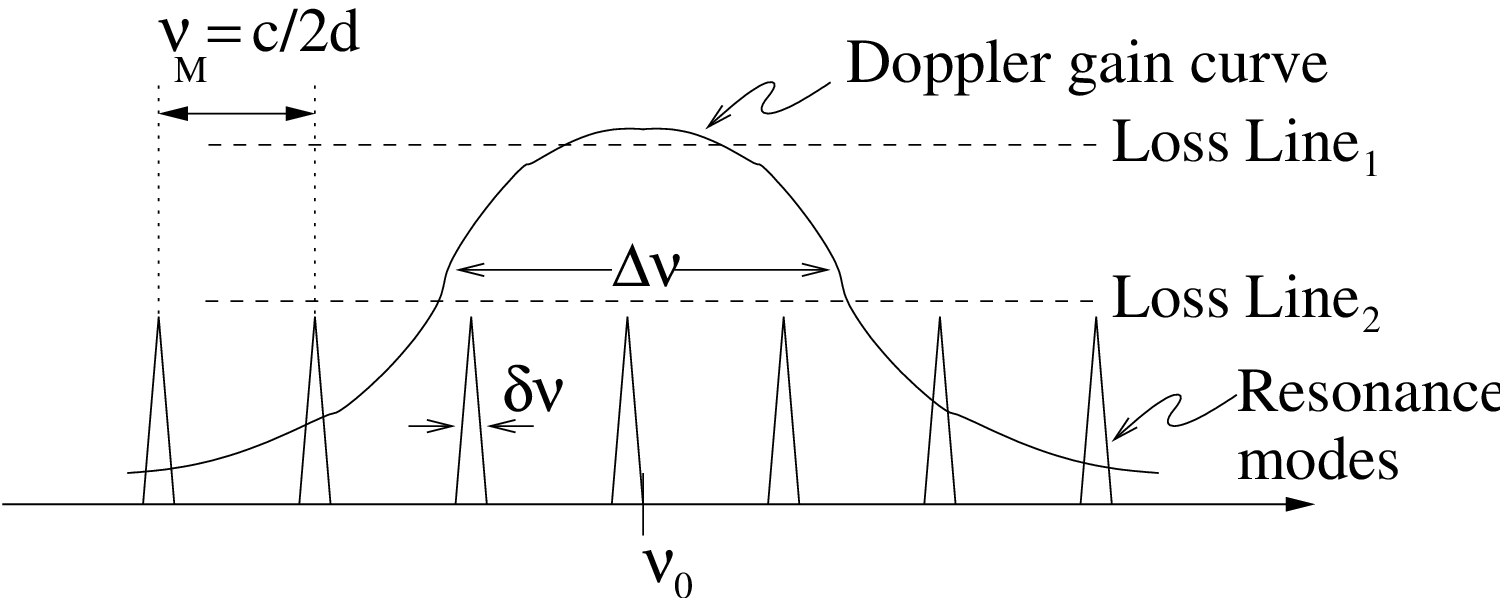}
\caption{\label{fig:laser-power-spectrum}Power spectrum of a laser showing Doppler gain curve, discrete frequency resonance modes and two possible levels of cavity loss\cite{Bridges::1968}.}
\end{figure}

The active medium between the mirrors can be considered as a narrow-band optical amplifier. The gain curve for this amplifier is centered around the frequency $\nu_{\circ}$, such that $h\nu_{\circ}$ is the energy released by the atomic transition that emits the photon. Only a few of the discrete frequencies given by Eq.\ref{eq:resonator-discrete-frequencies} appear in the laser beam. These are the amplified modes, the others are attenuated by the medium. For example, in a commercial red He-Ne laser, photons are emitted because of transition of Ne atoms from a $2p^55s$ state to $2p^53p$ state, which corresponds to a center frequency of $\nu_{\circ}\sim 473$ THz. The gain curve is primarily Doppler broadened\cite{Saleh::1991} by an amount
\begin{equation}
\Delta \nu \sim \frac{\nu_{\circ}}{c}\sqrt{\frac{2kT}{M}},
\label{eq:laser-doppler-width}
\end{equation}
where $k$ is Boltzmann's constant, $T$ is the gas temperature and $M$ is the molecular mass of the radiating atom. For a collection of Ne atoms emitting light at the He-Ne wavelength of $\lambda_{\circ}=632.8$ nm at room temperature, the Doppler width is $\sim2$ GHz. Thus, an amplifier with gain width $\Delta \nu  \sim 2$ GHz allows about $4$ modes separated by $\nu_M \sim 600$ MHz, as sketched in Fig.\ref{fig:laser-power-spectrum}.

Power will build up from noise in modes for which the gain exceeds the losses. As the power in modes builds up, modes will saturate and equilibrate, so that the gain balances the losses. Modes for which the losses exceed the gain are severely attenuated. For example, if loss-line $1$ in Fig.\ref{fig:laser-power-spectrum} represents the losses in the system, a monochromatic wave will exist, the one closest to the peak of the amplifier gain function. On the other hand, if the losses are represented by loss line $2$, there will be $3$ distinct modes in the wave. Power in various modes is distributed according to the amplifier gain profile and losses in the system\cite{Seigman::1986}.

The wave electric field in the polarization direction for an ideal laser can be represented as:
\begin{subequations}
\label{eq:resonator-electric-field}
\begin{eqnarray}
E(t) &=& \frac{1}{2\pi}\sum_q \tilde{E}_q e^{\ui\omega_q t} \\ \label{eq:resonator-electric-field-a}
              &=& \frac{1}{2\pi}\int\limits_{-\infty}^{\infty}\sum_q \tilde{E}_q \delta(\omega -\omega_q)e^{\ui\omega t}\ud \omega,  \label{eq:resonator-electric-field-b}
\end{eqnarray}
\end{subequations}
where $\omega_q=2\pi\nu_q$. Functions and variables in the frequency domain will be represented by a ``tilde.'' Equation \ref{eq:resonator-electric-field-b} is just a Fourier transform relation. Thus the Fourier transform of the electric field for an ideal laser is a series of delta functions, with the non-zero Fourier coefficients $\tilde{E}_q$ corresponding to the non-attenuated modes.

The discrete resonant frequency modes of a laser are each broadened by a small amount $\delta \nu$, due to:
\begin{enumerate}
\item Losses due to absorption and scattering within the medium\cite{Saleh::1991}. These losses relate to the finite photon decay time via the uncertainty relation between time and frequency.
\item Imperfect reflection at the mirrors\cite{Saleh::1991}.
\item Vibration of mirrors\cite{Collier::1971}. If the mirrors vibrate by an amount $\delta d$, the corresponding broadening of the modes is given by, $\delta \nu \sim \nu_{\circ}\delta d/d$.
\end{enumerate}
In most commercial lasers, the frequency broadening thus produced is of the order of $\delta \nu \sim 1$ MHz, as sketched in Fig.\ref{fig:laser-power-spectrum}. Typically, $\delta \nu \ll \nu_M$, and so the frequency-broadened modes do not overlap each other.

The Fourier transform of the electric field will now consist of a series of broadened functions. Under the simplifying assumption that all the modes are broadened by the same amount, the electric field Fourier transform can be represented as
\begin{equation}
\tilde{E}(\omega) = \sum_q \tilde{E}_q \tilde{F}(\omega - \omega_q),
\label{eq:resonator-electric-field-fourier-transform}
\end{equation}
where $\tilde{F}(\omega)$ is a low-pass broadening function of width $2 \pi \delta \nu$. Because the modes are well separated, the spectral power is
\begin{equation}
|\tilde{E}(\omega)|^2 = \sum_q |\tilde{E}_q|^2 |\tilde{F}(\omega - \omega_q)|^2.
\label{eq:laser-electric-field-power-spectrum}
\end{equation}
\subsection{\label{subsec:auto-correlation-to-power-spectrum}Phase auto-correlation function related to power spectrum}
The auto-correlation function of a laser is defined as
\begin{equation}
G(\tau)=\frac{\left\langle E^*(t)E(t+\tau)\right\rangle}{\left\langle|E(t)|^2\right\rangle}.
\label{eq:definition-auto-correlation}
\end{equation}
The coherence time of a laser is defined as the time $\tau$ at which the auto-correlation function $G(\tau)$ falls significantly below $1$ and the coherence length of a laser is the coherence time scaled by $c$. It is traditionally assumed that if an interferometer is set up with path length difference greater than the coherence length, the phases of the two waves will be uncorrelated, so no interference pattern will be observed. However, this standard concept of coherence length is misleading because the phase auto-correlation function is an almost periodic function; for the purpose of interferometry, it is sufficient to maintain a path length difference corresponding to a maximum of the auto-correlation function.

Using
\begin{equation}
E(t)=\frac{1}{2\pi}\int\limits_{-\infty}^{\infty}\tilde{E}(\omega)e^{\ui \omega t}\ud\omega,
\label{eq:fourier-transform-definition}
\end{equation}
Eq.\ref{eq:definition-auto-correlation} can be expressed as
\begin{equation}
G(\tau) = \frac{\int\limits_{-\infty}^{\infty} \int\limits_{-\infty}^{\infty} \ud \omega \, \ud \omega' \tilde{E}^*(\omega)\tilde{E}(\omega') e^{\ui \omega' \tau}\langle e^{\ui(\omega'-\omega)t}   \rangle}{\int\limits_{-\infty}^{\infty} \int\limits_{-\infty}^{\infty} \ud \omega \, \ud \omega' \tilde{E}^*(\omega)\tilde{E}(\omega') \langle e^{\ui(\omega'-\omega)t}   \rangle}.
\label{eq:simplified-auto-correlation-a}
\end{equation}
Using the relation
\begin{equation}
\langle e^{\ui (\omega' - \omega)t} \rangle \sim \int\limits_{-\infty}^{\infty}e^{\ui(\omega'-\omega)t}\ud t \sim \delta(\omega'-\omega),
\label{eq:delta-function-representation} 
\end{equation}
Eq.\ref{eq:simplified-auto-correlation-a} reduces to
\begin{equation}
 G(\tau) = \frac{\int\limits_{-\infty}^{\infty} \ud \omega |\tilde{E}(\omega)|^2 e^{\ui \omega \tau}}{\int\limits_{-\infty}^{\infty} \ud \omega |\tilde{E}(\omega)|^2},
\label{eq:simplified-auto-correlation-b}
\end{equation}
so
\begin{equation}
G(\tau)=\frac{1}{2\pi}\int\limits_{-\infty}^{\infty}\tilde{S}(\omega)e^{\ui \omega \tau} \ud \omega,
\label{eq:Wiener-Khinchin}
\end{equation}
where $\tilde{S}(\omega)$ is the normalized spectral power defined by
\begin{equation}
\tilde{S}(\omega) = \frac{2\pi|\tilde{E}(\omega)|^2}{\int\limits_{-\infty}^{\infty} \ud \omega |\tilde{E}(\omega)|^2}.
\label{eq:definition-normalized-spectral-power}
\end{equation}
Equation \ref{eq:Wiener-Khinchin} is the Wiener-Khinchin theorem\cite{Saleh::1991} and shows that the auto-correlation function and the normalized spectral power are Fourier transform pairs.

Using Eq.\ref{eq:laser-electric-field-power-spectrum}, the auto-correlation function has the dependence
\begin{equation}
G(\tau) \sim \int\limits_{-\infty}^{\infty} \ud \omega \sum_q|\tilde{E}_q|^2 |\tilde{F}(\omega-\omega_q)|^2e^{\ui \omega \tau}.
\label{eq:auto-correlation-manipulation-a}
\end{equation}
Let
\begin{equation}
\tilde{\mathcal{F}}(\omega)=|\tilde{F}(\omega)|^2
\label{eq:script-tilde-F-definition}
\end{equation}
and let $\mathcal{F}(\tau)$ be the Fourier inverse of $\tilde{\mathcal{F}}(\omega)$ so
\begin{equation}
\tilde{\mathcal{F}}(\omega)=\int\limits_{-\infty}^{\infty}\mathcal{F}(\tau)e^{-\ui \omega \tau}\ud \tau.
\label{eq:script-F-fourier-inverse}
\end{equation}
Since $\tilde{\mathcal{F}}(\omega)$ has a spread of $\sim 2 \pi \delta \nu$, $\mathcal{F}(\tau)$ will have a spread of $\sim 1/\delta \nu$.

From Eq.\ref{eq:auto-correlation-manipulation-a},
\begin{eqnarray}
G(\tau) &\sim& \int\limits_{-\infty}^{\infty} \ud \omega \sum_q|\tilde{E}_q|^2 \left( \int\limits_{-\infty}^{\infty}\mathcal{F}(\tau')e^{-\ui (\omega-\omega_q)\tau'}\ud\tau'\right)e^{\ui \omega \tau} \nonumber \\
        & = & \int\limits_{-\infty}^{\infty} \ud\tau'\sum_q|\tilde{E}_q|^2 e^{+\ui \omega_q \tau'} \mathcal{F}(\tau')\left( \int\limits_{-\infty}^{\infty}e^{-\ui \omega(\tau'- \tau)}\ud \omega\right) \nonumber \\
        & \sim & \int\limits_{-\infty}^{\infty} \ud\tau'\sum_q|\tilde{E}_q|^2 e^{\ui \omega_q \tau'} \mathcal{F}(\tau')\delta(\tau'- \tau) \nonumber \\
        & =    & \sum_q|\tilde{E}_q|^2 e^{\ui \omega_q \tau} \mathcal{F}( \tau ) \nonumber \\
        & =    &  \mathcal{P}(\tau) \mathcal{F}( \tau ),
\label{eq:auto-correlation-manipulation-main}
\end{eqnarray}
where 
\begin{eqnarray}
\mathcal{P}(\tau)&=&\sum_q|\tilde{E}_q|^2 e^{\ui \omega_q \tau} \nonumber \\
                 &=&\sum_q|\tilde{E}_q|^2e^{\ui2\pi q\nu_M\tau}.
\label{eq:definition-script-P}
\end{eqnarray}
Each of the complex exponentials in Eq.\ref{eq:definition-script-P} is periodic in $\tau$, with a period of $\frac{1}{\nu_M}$. Thus, $\mathcal{P}(\tau)$ is also periodic with the same period. For $\tau=0,\frac{1}{\nu_M},\frac{2}{\nu_M},\frac{3}{\nu_M},\cdots$, all the components add up constructively and so $\mathcal{P}(\tau)$ will be maximum at these values of $\tau$. The exact shape of $\mathcal{P}(\tau)$ will depend on the value of the coefficients $|\tilde{E}_q|^2$. For a laser with large number of modes (corresponding to many non-zero $|\tilde{E}_q|^2$'s), $\mathcal{P}(\tau)$ may have a steep decay away from its peaks.

For a typical laser with $d\sim25$ cm and $\delta\nu\sim 1$ MHz, $\mathcal{P}(\tau)$ will be periodic with period $1.67$ ns and $\mathcal{F}(\tau)$ will have a spread of $1\ \mu$s. It is convenient to scale time with $c$ to express $\mathcal{P}$, $\mathcal{F}$ and $G$ as functions of length. $\mathcal{P}(\delta L)$ is thus periodic with period $2d=0.5$ m and $\mathcal{F}(\delta L)$ decreases with a $300$ m scale length. $G(\delta L)$ is thus the product of a slowly decaying envelope function $\mathcal{F}(\delta L)$ and a periodic function $\mathcal{P}(\delta L)$. The interferometer at Caltech operates at $\delta L \sim 8$ m corresponding to the $16^{th}$ maximum of $\mathcal{P}(\delta L)$. We assume $\mathcal{F}(\delta L)$ to be Gaussian $\sim e^{-{(\delta L)^2}/{2l^2}}$, where $l$ the width of $\mathcal{F}$, is approximately $300$ m. If an interferometer is operated at a path length difference corresponding to a maximum of $\mathcal{P}(\delta L)$, the strength of the interference signal will be proportional to $\mathcal{F}(\delta L)$. Thus for the Caltech interferometer, operating at a path length difference of $8$ m causes attenuation of the signal amplitude by a factor $\mathcal{F}(\delta l = 8\textrm{ m})=0.9996$. Or in other words, only $0.04\%$ of power is lost due to unequal path length effects and phase coherence is maintained since $\mathcal{P}(\delta l)$ has the same value at $8$ m as at $0$ m.
\subsection{\label{subsec:measure-auto-correlation}Measurement of laser phase auto-correlation function}
\begin{figure}[p]
\includegraphics[width=0.45\textwidth]{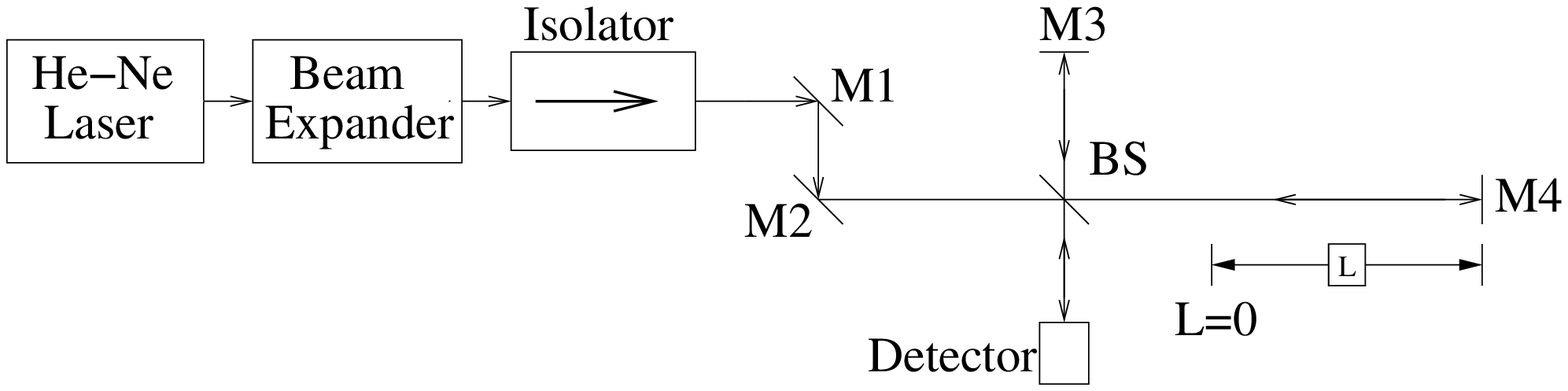}
\caption{\label{fig:APh24-setup}Michelson setup to measure phase auto-correlation of laser. BS stands for beam splitter and M for mirror}
\end{figure}
\begin{figure}[p]
\includegraphics[width=0.45\textwidth]{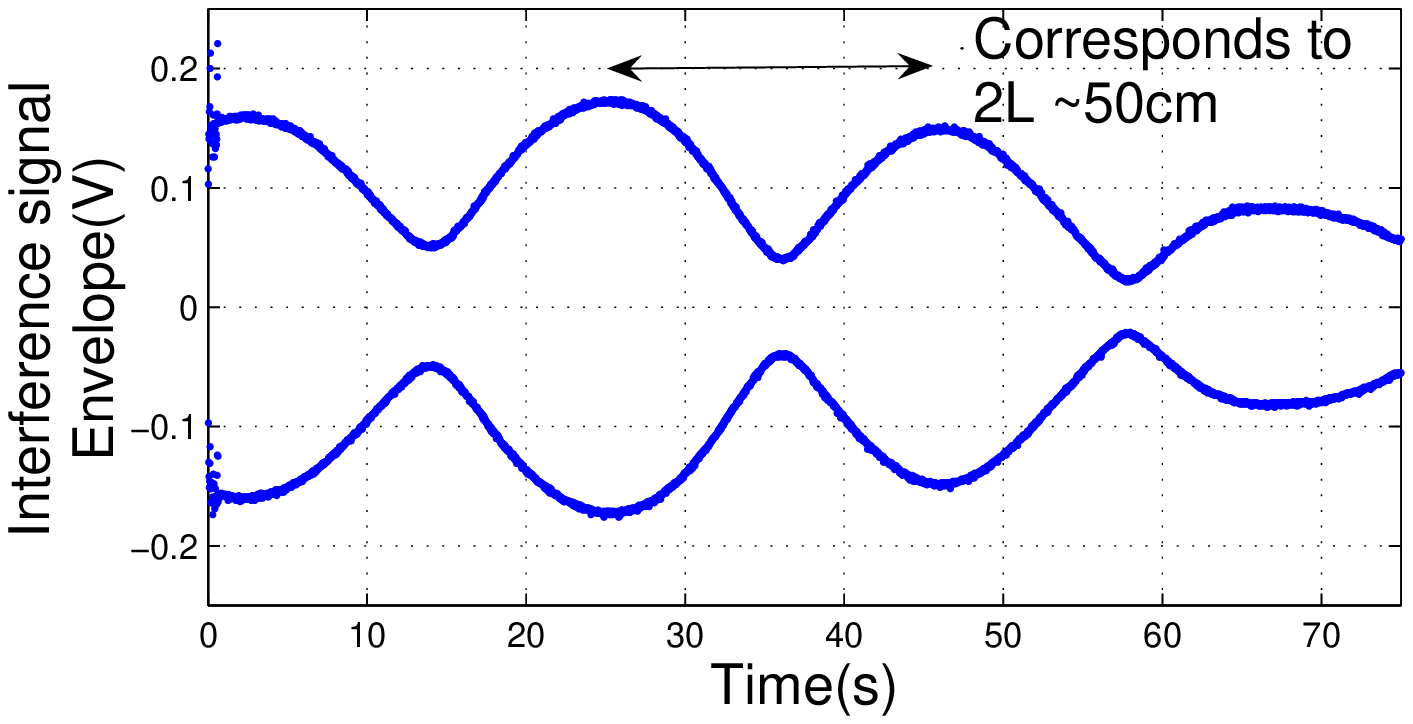}
\caption{\label{fig:laser-APh24-signal}Envelope of the interference signal measured using the setup shown in Fig.\ref{fig:APh24-setup}. The path length difference $2L$ was varied at a constant rate. The envelope magnitude is directly proportional to the phase auto-correlation function.}
\end{figure}
To test if the laser being used in the Caltech interferometer indeed has a periodic auto-correlation function, the laser was used in the Michelson interferometer setup shown in Fig.\ref{fig:APh24-setup}. Mirror M4's location, $L$, was varied with a linear translation stage at a constant speed and the amplitude of the interference signal was plotted as a function of time, as shown in Fig.\ref{fig:laser-APh24-signal}. Interference is caused by ambient noise vibrating the mirrors. At $t=0$ s, $L$ was $0$, and thus the path lengths were approximately equal. The amplitude of the interference signal is directly proportional to the phase auto-correlation function. Since $L$ was increased at a constant rate, the horizontal time axis in Fig.\ref{fig:laser-APh24-signal} is proportional to the path length difference $2L$. As seen from Fig.\ref{fig:laser-APh24-signal}, the phase auto-correlation function is periodic. The difference between successive maxima corresponded to $2L\sim50$ cm. The amplitude decreased significantly beyond the third maximum because the interferometer became misaligned with large motions of the mirror. The minima of the amplitude of the interference signal was of the order of the noise level of the detector.
\section{\label{sec:interferometer-setup}Interferometer Setup}
\begin{figure}[p]
\includegraphics[width=0.45\textwidth]{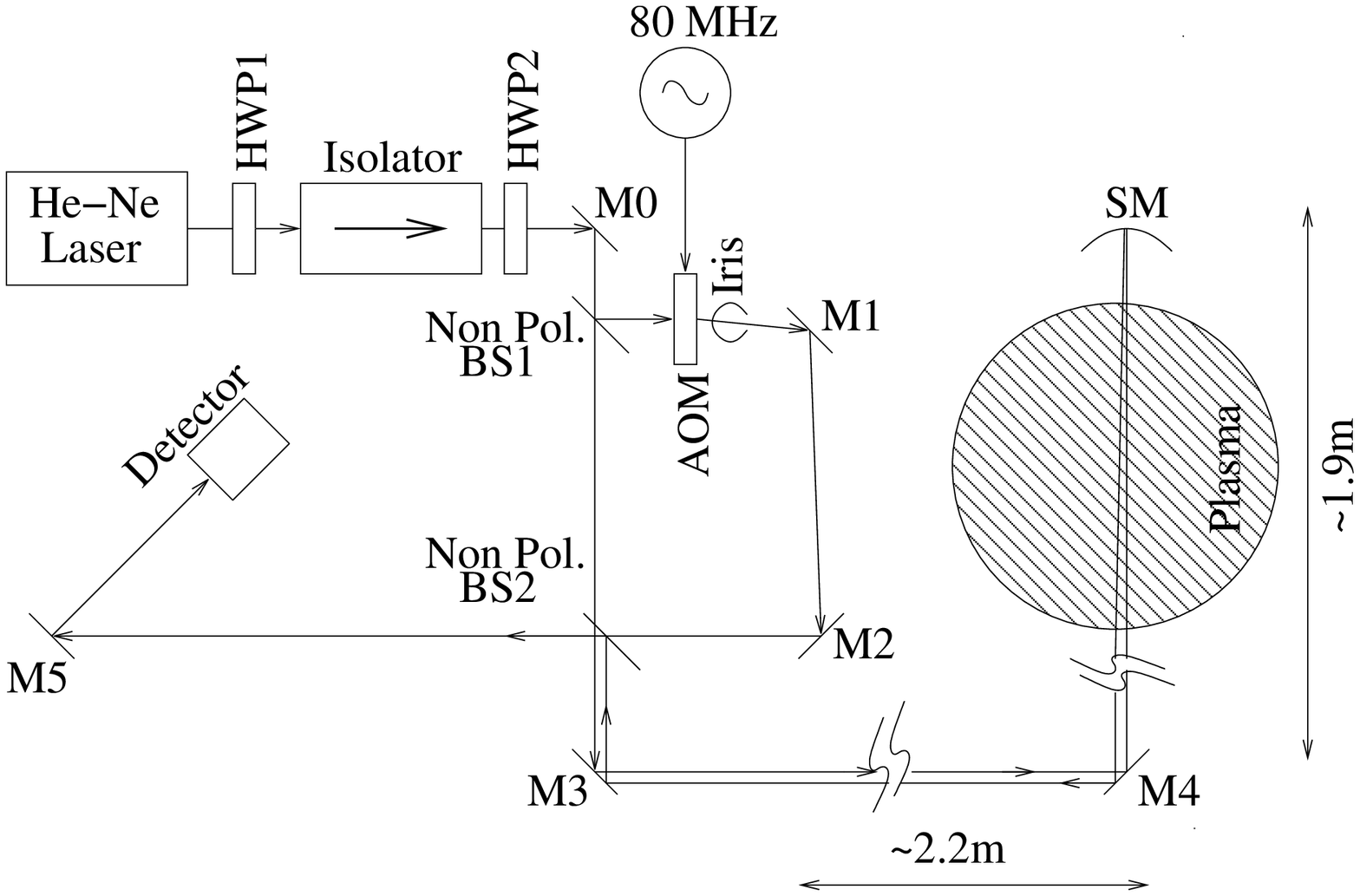}
\caption{\label{fig:heterodyne}Setup of the heterodyne interferometer for the Caltech spheromak formation experiment.}
\end{figure}
\begin{figure}[p]
\includegraphics[width=0.45\textwidth]{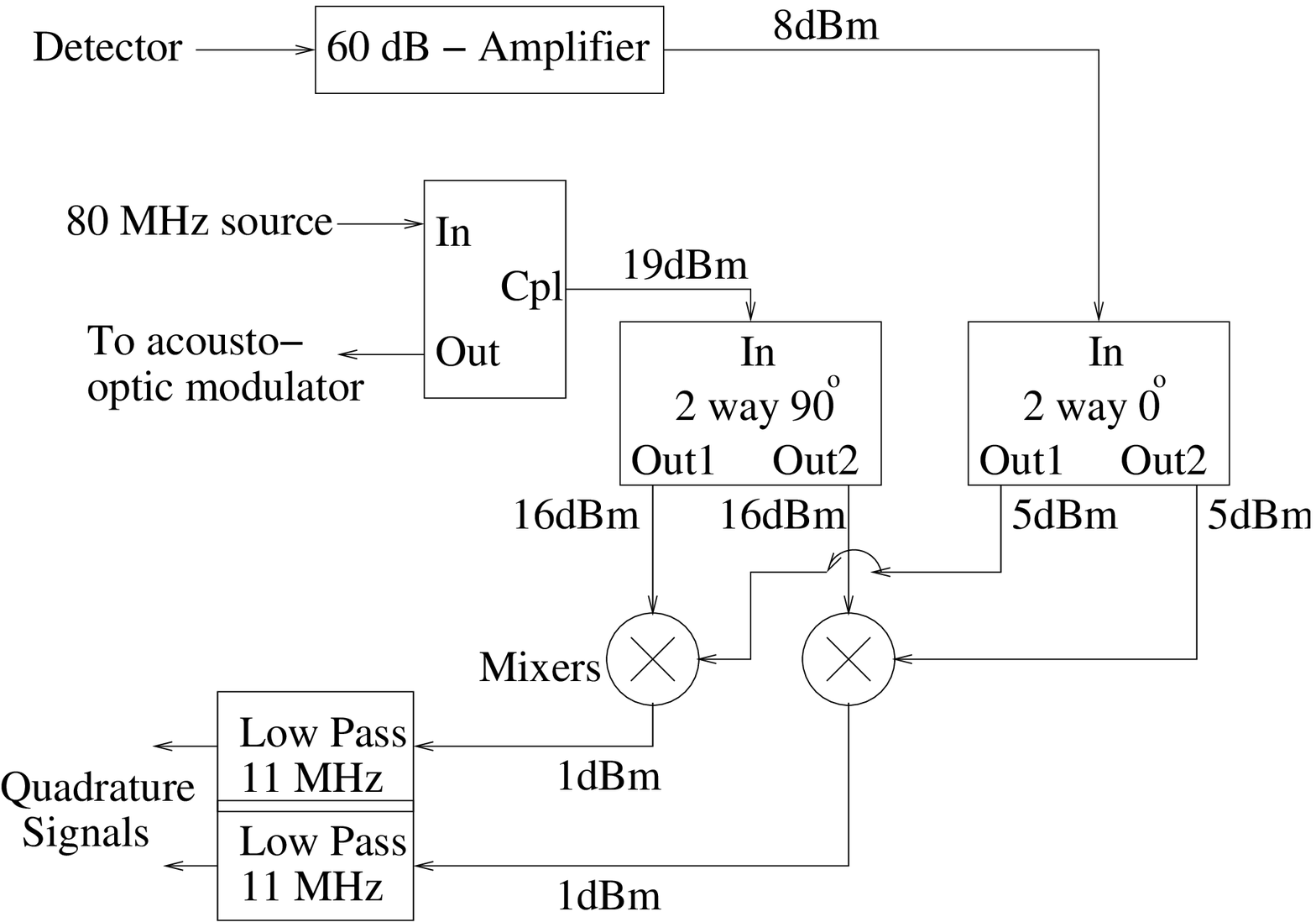}
\caption{\label{fig:heterodyne-electrical}RF circuit for the heterodyne interferometer. Typical signal power in dBm is mentioned for each connection.}
\end{figure}
Figure \ref{fig:heterodyne} shows the schematic for the interferometer. RF electronics for demodulating the signal is shown in Fig.\ref{fig:heterodyne-electrical}. Many interferometers, used on existing plasma experiments are two-color interferometers\cite{Baker:D3D:1978,Carlstrom:D3D:1988,Kawano::1992,Yasunori:JT60U:1997,Irby::1999,Acedo:TJ2:2000,Irby:TJ2:2001,Innocente:MST:1992,Acedo:TJ2:2004}, that decouple the phase shift caused by the plasma and by mechanical vibrations. Because mechanical vibrations are unimportant for the fast time scales($\sim 10$ $\mu$s) of the Caltech plasma experiment, a single laser interferometer is adequate.

The interferometer described in Fig.\ref{fig:heterodyne} is set up in a double pass geometry. By interfering beams with large path length difference, it was possible to locate most of the optical components on a small and accessible optical bench($18''\times 18''$). Mirror M4 and the spherical mirror SM are mounted on the vacuum chamber and direct the laser beam through the plasma and back to the optical bench. The design uses a $4$ mW linearly polarized He-Ne laser with a cavity length of $25$ cm. An optical isolator is placed in front of the laser to prevent any reflected laser light from entering the laser. Half wave plate HWP1 rotates the polarization vector of the laser beam so that it aligns with the direction of the polarizer at the input of the isolator. Half wave plate HWP2 transforms the beam coming out of the isolator into a vertically polarized beam. The polarization of the vertically polarized reference beam is unaltered upon reflection from mirrors, beamsplitter or from transmission through the acousto-optic modulator (AOM). Mirror M4 and the spherical mirror SM direct the beam into the vacuum chamber through sapphire windows. Sapphire is a birefringent material, and the windows are oriented to minimize the change in the polarization of the scene beam. Note that the interference signal is maximized if the polarization of the interfering beams is the same.

The AOM is oriented so that most of the power of the input reference beam is coupled into the first harmonic. Up to $86\%$ of the input power can be coupled into the first harmonic. The iris obstructs all other beams except for the first harmonic.

The radius of curvature of the spherical mirror is $4$ m, the approximate distance the beam travels from the optical table to the spherical mirror, so the spherical mirror focuses the beam back to almost its original size. The spherical mirror position is adjusted to ensure that the path length difference between the scene and reference beams is approximately an even multiple of the laser cavity length. The path length difference is changed in steps of $\sim15$ cm. From Fig.\ref{fig:laser-APh24-signal}, it is observed that the laser phase auto-correlation function is flat around the maximas and thus precise adjustment of the path lengths is unnecessary. Once the path length difference is maintained in the proximity of a maxima, the signal amplitude of the interferometer was found to depend more on the alignment of the scene and reference beams than on the precise path length difference of the two beams.

A major advantage of the design shown in Fig.\ref{fig:heterodyne} is the ease of alignment. The two beams are arranged to overlap each other simply by adjusting the cube beam splitter BS2, and the mirror M2. Both these components are located on the optical bench and are easily accessible. A cube beam splitter was used for combining the beams instead of a plate beam splitter since a cube beam splitter does not introduce any lateral shift in the position of the beam passing through.
\section{\label{sec:results}Results}
\begin{figure}[p]
\includegraphics[width=0.45\textwidth]{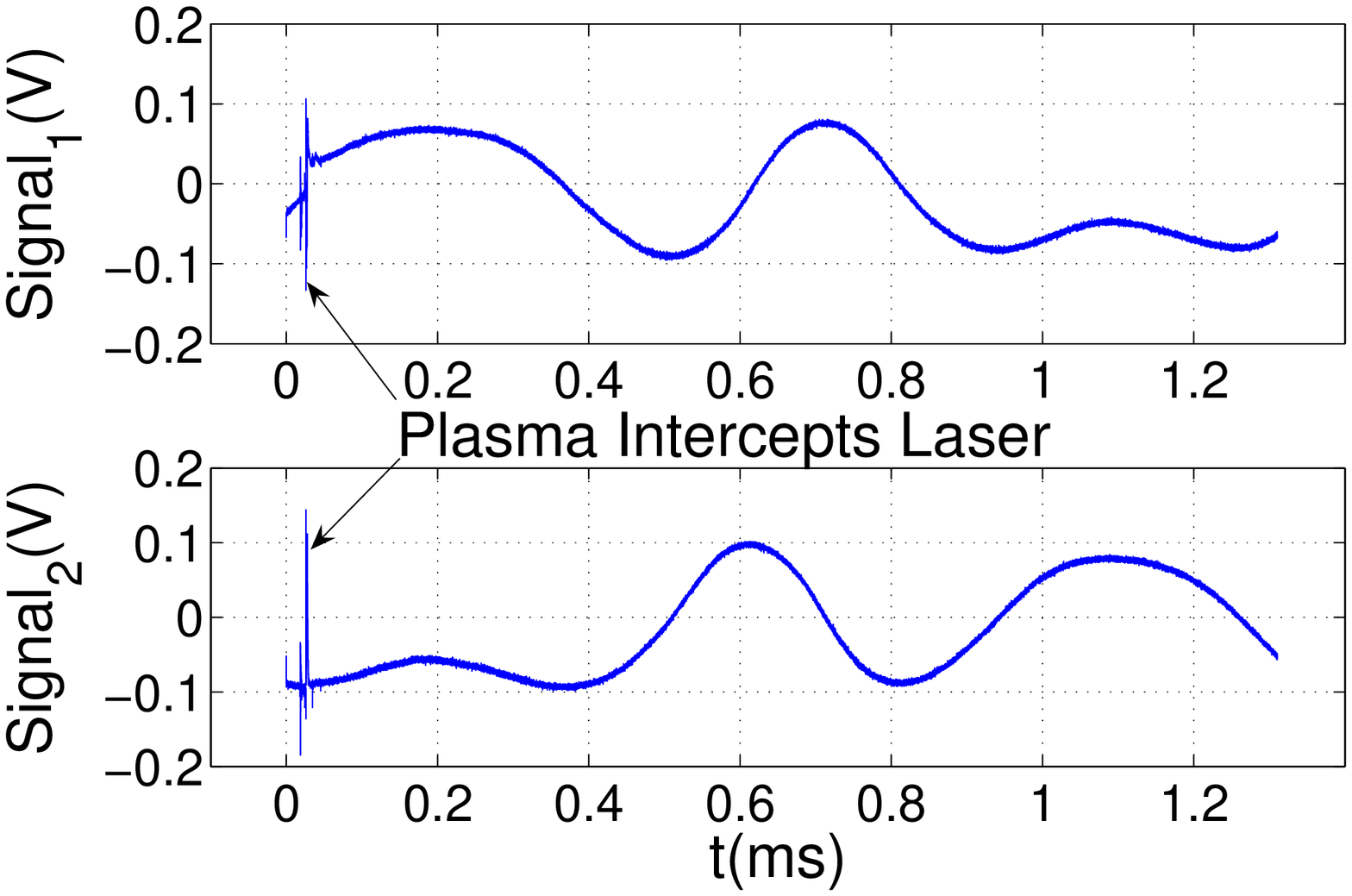} 
\caption{\label{fig:heterodyne-quadrature-signals}Quadrature signals recorded by the digitizer from shot\#7466.}
\end{figure}
\begin{figure}[p]
\includegraphics[width=0.45\textwidth]{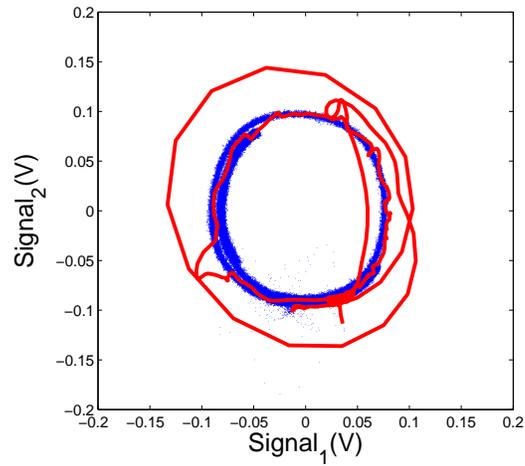}
\caption{\label{fig:heterodyne-quadrature-signals-xy-plot}(Color online) Lissajous figure of the data shown in Fig.\ref{fig:heterodyne-quadrature-signals}. The data set corresponding to plasma intercepting the laser beam is plotted as a solid red line while the non-plasma times are plotted in blue dots.}
\end{figure}
\begin{figure}[p]
\includegraphics[angle=-90,width=.45\textwidth]{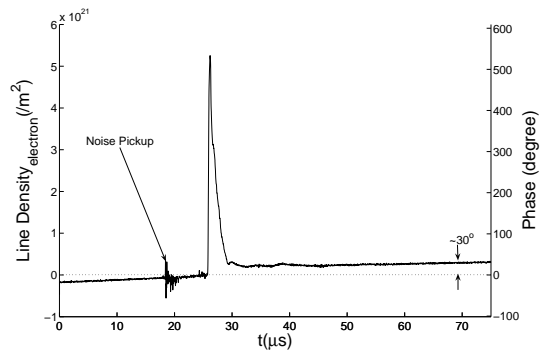} 
\caption{\label{fig:heterodyne-density-signal}Line average density interpreted from the signals shown in Fig.\ref{fig:heterodyne-quadrature-signals}. The axis for the plasma induced phase change is shown on the right. Note the noise pickup from the discharge of capacitor banks.}
\end{figure}
\begin{figure}[p]
\includegraphics[width=.45\textwidth]{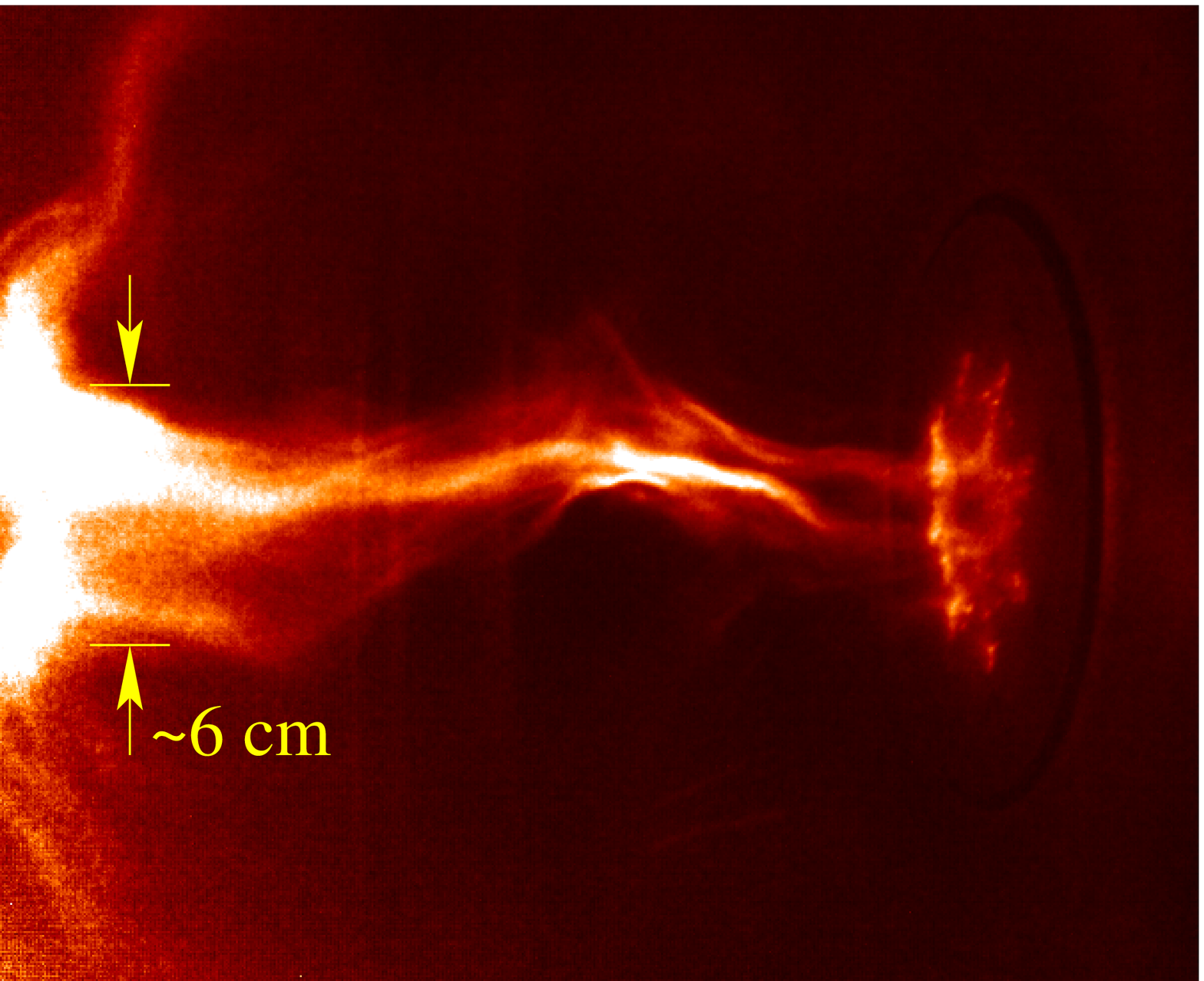} 
\caption{\label{fig:imacon}(Color online) Image of the plasma from shot\#7466, frame 16. The laser beam passes through a plasma column approximately $6$ cm in diameter. As the interferometer is designed in a double pass geometry, the length of plasma, the laser beam traverses is $L \sim 12$ cm.}
\end{figure}

Figure \ref{fig:heterodyne-quadrature-signals} shows the demodulated quadrature signals from a typical plasma shot. Note that when one of the signals is at its maximum or minimum, the other is passing through zero - a consequence of being in quadrature. The random variation in the signal is due to acoustic frequency mechanical vibrations. Plasma causes the sudden change in the signal as indicated by the arrows.

The two signals shown in Fig.\ref{fig:heterodyne-quadrature-signals} are plotted against each other in Fig.\ref{fig:heterodyne-quadrature-signals-xy-plot}. The data set corresponding to plasma intercepting the laser beam is plotted as a solid red line while the non-plasma times are plotted in blue dots. The extent to which the signals are in quadrature can be estimated from the extent to which Fig.\ref{fig:heterodyne-quadrature-signals-xy-plot} resembles a circle. Note that refractive bending intensified the signal amplitude when the plasma intercepts the beam. Provided that the beams undergo only a ``small'' displacement because of refractive bending, taking the ratio of the two signals removes the effects of refractive bending on the phase inferred\cite{Lowenthal::1979}.

Taking the inverse tangent of the ratio of the two quadrature signals yields the phase change caused by the plasma. The plasma density shown in Fig.\ref{fig:heterodyne-density-signal} was estimated from the phase change using
\begin{equation}
\int\limits_{0}^{L}n(x) \ud x = \frac{4 \pi c^2 m_e \epsilon_{\circ}}{e^2 \lambda_{\circ}}\Delta \phi_p,
\label{eq:density-from-phase}
\end{equation}
where $\Delta \phi_p$ is the phase shift induced by the plasma, $m_e$ is the mass of electron, $\epsilon_{\circ}$ is the permittivity of free space, $e$ is the electron charge, $L$ is the length of plasma the laser beam traverses, and $\lambda_{\circ}=632.8$ nm is the He-Ne laser wavelength. From Fig.\ref{fig:heterodyne-density-signal} it is observed that the mechanical vibrations have changed the phase by $\sim30^{\circ}$ in $\sim50\ \mu s$, while the plasma has induced a phase change of $>500^{\circ}$ and back in $\sim15\ \mu s$. Thus the mechanical vibrations have negligible effect on inferred plasma density, and can even be filtered out during post processing of data.

Line-densities of the order of $5 \times 10^{21}/\textrm{m}^2$ were observed in the experiment. Assuming a double pass plasma length $L\sim12$ cm, as shown in Fig.\ref{fig:imacon}, corresponds to average densities of $\sim 4 \times 10^{22}/\textrm{m}^3$. These results are in good agreement with the density inferred by fitting the Stark broadened $H_\beta(486.133\ \textrm{nm})$ spectral line profile to a Lorentzian\cite{Yun:caltech:2006}. The signal to noise performance of interferometers is specified in terms of the phase ambiguity of the signals. The phase ambiguity is given by $\sigma/A$, where $\sigma$ is the rms error in the signal and $A$ is the strength of the signal\cite{Buchenauer:LANL:1977}. For typical data this was found to be $\sim 1 ^{\circ}$, corresponding to a line-density error of $\sim 10^{19}/\textrm{m}^2$.

\section{\label{sec:conclusion}Conclusion}
A He-Ne heterodyne interferometer was developed for the Caltech spheromak formation experiment. The design is especially suited for fast plasma experiments with time scales much smaller than the time scales of mechanical vibrations of the mirrors. The interferometer operates well even though there is a path length difference of $\sim 8$ m between the scene and the reference beams. Operating at such a large path length difference considerably reduced the number of optical components and difficulties in alignment.

The laser beams in a two-color interferometer may have different periods for their phase auto-correlation function. Thus operating at large path length differences in a two-color interferometer may require some non-trivial modifications to existing setups. However, for two-color interferometers with lasers having the same cavity length, or using a single laser\cite{Irby::1999}, the path length of the reference beam could be adjusted straightforwardly.
\begin{acknowledgments}
Helpful discussions with Prof. William B. Bridges and Prof. Kerry Vahala on laser phase auto-correlation are gratefully acknowledged. We would also like to thank Dr. Heun-Jin Lee for his technical help in setting up the Michelson interferometer in Fig.\ref{fig:APh24-setup} and Gunsu Yun for spectroscopic measurements. In addition, we would like to recognize the helpful initial guidance from Dr. Raymond P. Golingo, University of Washington, Seattle, in designing the heterodyne interferometer.

This work was supported by USDOE Grant DE-FG02-04ER54755.
\end{acknowledgments}

\end{document}